# Thermodynamics of nanodomain formation and breakdown in Scanning Probe Microscopy: Landau-Ginzburg-Devonshire approach

Anna N. Morozovska,[*, 1] Eugene A. Eliseev,[2] Sergei V. Svechnikov,[1] Peter Maksymovych,[3] and Sergei V. Kalinin[†, 3]

[1] Institute of Semiconductor Physics, 41, pr. Nauki, 03028 Kiev, and
[2] Institute for Problems of Materials Science, 3, Krjijanovskogo, 03142 Kiev,
National Academy of Sciences of Ukraine, Ukraine
[3] Oak Ridge National Laboratory, Oak Ridge, TN 37831, USA

**Abstract**

Thermodynamics of tip-induced nanodomain formation in scanning probe microscopy of ferroelectric films and crystals is studied using the Landau-Ginzburg-Devonshire phenomenological approach. The local redistribution of polarization induced by the biased probe apex is analyzed including the effects of polarization gradients, field dependence of dielectric properties, intrinsic domain wall width, and film thickness. The polarization distribution inside "subcritical" nucleus of the domain preceding the nucleation event is very smooth and localized below the probe, and the electrostatic field distribution is dominated by the tip. In contrast, polarization distribution inside the stable domain is rectangular-like, and the associated electrostatic fields clearly illustrate the presence of tip-induced and depolarization field components. The calculated coercive biases of domain formation are in a good agreement with available experimental results for typical ferroelectric materials. The microscopic origin of the observed domain tip elongation in the region where the probe electric field is much smaller than the intrinsic coercive field is the positive depolarization field in front of the moving counter domain wall. For infinitely thin domain walls local domain breakdown through the sample depth appears. The results obtained here are complementary to the Landauer-Molotskii energetic approach.

---

[*] Corresponding author: morozo@i.com.ua
[†] Corresponding author: sergei2@ornl.gov

## 1. Introduction
### 1.1. Local bias-induced phase transitions by SPM

Bias-induced phase transitions and order-parameter dynamics in polar materials are a subject of substantial experimental and theoretical interest. The examples include polarization switching in ferroelectric materials with applications to information storage and memory technologies [1, 2, 3], antiferroelectric-ferroelectric phase transitions and energy storage [4], and a broad gamut of bias-induced transitions between ergodic, non-ergodic, and ferroelectric states in ferroelectric relaxors [5]. Traditionally, these phenomena are studied macroscopically using the variants of capacitance and current detection techniques [6] or interferometric detection [7, 8, 9, 10]. In these studies, the information on local mechanisms controlling the nucleation and initial stages of phase transformation is essentially lost and only averaged distributions of switching parameters and activation energies can be extracted [11, 12, 13]. This limitation is common for all polar materials with reversible bias induced transitions, and extends to other systems with partially reversible and irreversible transitions, including phase-change materials [14], electrochemical [15] and solid-state reactions [16].

The emergence of the Scanning Probe Microscopy (SPM) based techniques in the last decade opens the way to concentrate electric field within a nanoscale volume of material [17, 18], thus inducing a local phase transition. This *field-localization* approach is complementary to a classical approach in nanoscience of *material-confinement* (e.g. using the nanoparticles, etc) and allows studying local properties avoiding the effect of surfaces and interfaces. For ferroelectric materials, the strongly inhomogeneous electric field causes polarization reversal in the nanosized region that can be used as a functional basis of data storage [19, 20] as well as a probing technique to study local mechanisms of domain nucleation, growth, and relaxation [21, 22, 23, 24, 25, 26].

In Piezoresponse Force Spectroscopy approach, local polarization switching is combined with the detection of electromechanical response [27] to yield the information on domain growth below the SPM tip [28]. Spatially resolved Piezoresponse Force Spectroscopy (PFS) was used to study polarization switching in the small volumes with negligible defect concentration [29], map distribution of random bond- and random field components of disorder potential [30], and map polarization switching on a single defect center [31].

These experimental developments have necessitated the theoretical analysis of domain nucleation mechanisms in the field of the SPM probe on the ideal surface [32, 33, 34 35, 36] and in the presence of charged defects [37, 38]. To date, the vast majority of these studies have been performed in the rigid dielectric approximation, as summarized below.

*1.2. Phenomenological approaches to nanoscale polarization reversal*

The approximation of "rigid ferroelectric" was originally used by Landauer [39] for the free energy calculations of the semi-ellipsoidal domain nucleation in a homogeneous electric field of a plain capacitor, and was later extended to predict the thickness dependence of the coercive field by Kay and Dunn [40]. A similar approximation was used by Miller and Weinreich [41] to study the domain wall motion, and extended by Sidorkin [42] to analyze the wall-defect interactions. Huber [43] considered the impact of the electromechanical coupling on the domain nucleation in the homogeneous external field. In this model, the domain walls between the regions with field-independent (i.e. "rigid") spontaneous polarization $\pm P_S$ are regarded as ultra-sharp (mathematically infinitely thin). The polarization adopts it bulk value within the domains and changes stepwise at the infinitely thin domain wall between them.

This approach was utilized in a series of works by Molotskii et al. [44, 32, 35] for the analysis of the domain formation caused by the inhomogeneous electric field of the biased SPM probe. The most striking result obtained by Molotskii et al is the "ferroelectric breakdown", namely the stable spike-like domain appearance with submicron radius $r$ and length $l$ of 10-100 microns, i.e. the polarization reversal appears in the spatial region, where the vanishing field of the probe is much smaller than intrinsic coercive field. Molotskii et al explained this behavior from the free energy consideration. Within the Landauer-Molotskii (LM) thermodynamic approach, the nucleus sizes and the equilibrium radius $r$ and length $l$ of semi-ellipsoidal domain are calculated from the free energy excess $G(V,r,l) = G_S(r,l) + G_V(V,r,l) + G_{DL}(r,l)$, where the positive domain wall surface energy $G_S(r,l) \sim \psi_S \, l \, r$ at $l >> r$ ($\psi_S$ is the surface energy density). The Landauer depolarization field energy $G_{DL}(r,l)$ is positive and proportional to $r^4/l$ at $l \gg r$, and so it vanishes as $1/l$ with the increase of the domain length. The negative probe field-domain interaction energy, $G_V(V,r,l) \sim -V r^2 l / \left(\sqrt{r^2+d^2}+d\right)\left(\gamma\sqrt{r^2+d^2}+\gamma d+l\right)$ ($\gamma$ is dielectric anisotropy factor, $V$ is applied bias), is proportional to $rl/(r+l)$ when the domain radius $r$ exceeds the characteristic size of the tip, $d$, and so it saturates with domain length increase. The condition of negligible surface energy ($\psi_S=0$) leads to the domain breakdown $l\to\infty$ even at infinitely small bias $V$.

This thermodynamic analysis was further developed by Morozovska et al. to account for the finite electric field below the probe, surface and bulk screening, etc [45, 36]. In particular, this analysis allows the description of bias-dependence of the saddle point on the free energy surface $G(V,r,l)$, i.e. the activation energy for nucleation. It was found that the activation energy is $\sim V^{-3}$, where $V$ is the applied bias, and in this model the nucleation process is thermally activated. For typical materials parameters, the corresponding activation energies are in the 0.1 – 10 eV range. However, recent experimental studies have illustrated that temperature dependence of activation bias is much weaker than predicted by the rigid model [46]; similarly, the comparison of the phase field modeling and experimental measurements indicates that the switching mechanisms in PFM is close to intrinsic [29].

### *1.3. Polarization switching in the LGD approximation*

The self-consistent description of the SPM probe-induced domain formation in the ferroelectrics and other ferroics requires an analytical approach based on the Landau-Ginzburg-Devonshire (LGD) thermodynamic theory. For ferroelectrics, LGD describes the dynamics of a continuous spatial distribution of the polarization vector **P** in an arbitrary electric field and the nonlinear long-range polarization interactions (correlation effects) [47]. In this manner, the LGD-approach avoids the typical limitations (sharp walls and field-independent polarization value) of the rigid ferroelectric approach [compare Fig. 1 (b) and 1 (c)]. Charge-neutral 180°-domain walls do not cause the depolarization electric field and usually are ultra-thin. However, the charged (or counter) domain wall at the domain apex creates a strong depolarization field due to uncompensated bound charges ($\text{div}\mathbf{P} \neq 0$). The charged wall inevitably appears at the tip of the nucleating domain [Fig.1 (b)].

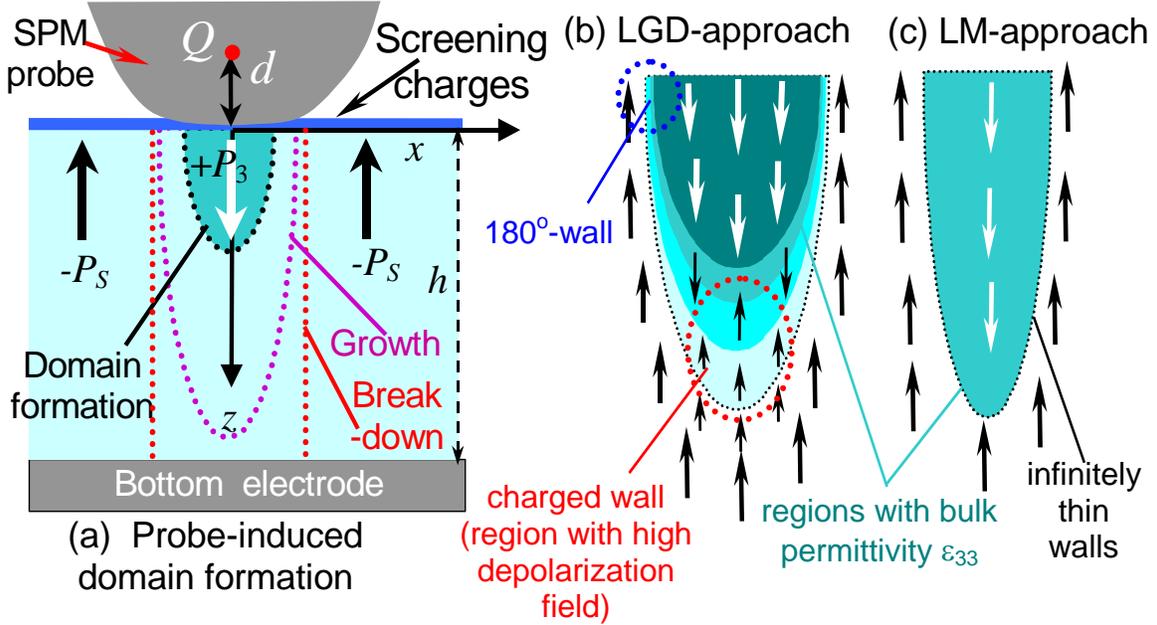

**FIG. 1**. (a) Schematics of the domain nucleation caused by the strongly inhomogeneous electric field of the biased SPM probe in contact with the sample surface. (b, c) Characteristic aspects of LGD-approach (b) and rigid LM-approach (c).

Polarization switching on the BiFeO$_3$ (100) ideal surface [29] and in the presence of the well-defined bicrystal grain boundary [48] was recently studied numerically using phase-field modeling. This analysis has confirmed the formation of a soft subcritical nucleus for the bias below nucleation. Above the nucleation threshold, the formation of needle-like domains as well as domain wall broadening at the domain apex has been observed. However, the limitations of the system size for the 3D phase field modeling preclude the analytical determination of the domain shape when the domain size significantly exceeds the tip size. Similarly, screening at the surface and the domain apex are difficult to access systematically.

Previously, the interaction of the ferroelectric 180°-domain wall with a strongly inhomogeneous electric field of the biased probe was studied *analytically* for a second order ferroelectric within the LGD-approach , Ref. [49]. The approximate analytical expressions for the equilibrium distribution of *surface* polarization were derived from the *free energy functional* by a direct variational method using the integration over a spatial region. However, *local* consideration of the electric field distribution, nonlinear and correlation effects is necessary for a reliable analysis of the polarization depth profile and the length of tip-induced domains in both first and second order ferroelectric materials.

In this paper we adopt the local LGD-based approach for the description of the polarization dynamics in the local electric field of an SPM probe. The analytical expressions for both first and second order ferroelectrics are derived. Both the pre-nucleation and post-nucleation stages are analyzed. This analysis provides insight into the effects of the intrinsic wall width, electrostatic potential distribution of the probe, ferroelectric material parameters and the nonlinear correlation and depolarization effectson the local polarization dynamics.

## 2. The problem statement

Here we study polarization switching in a uniaxial ferroelectric material. The spontaneous polarization $P_3(\mathbf{r})$ is directed along the polar axis, *z*. The sample is dielectrically isotropic in transverse directions, i.e. permittivities $\varepsilon_{11}$ and $\varepsilon_{22}$ are equal, while the $\varepsilon_{33}$ value may be different. The dependence of the in-plane polarization components on the electric field is linearized as $P_{1,2} \approx -\varepsilon_0(\varepsilon_{11}-1)\partial\varphi(\mathbf{r})/\partial x_{1,2}$. Then the problem for the electrostatic potential $\varphi(\mathbf{r})$ inside the material acquires the form:

$$\begin{cases} \varepsilon_{33}^b \frac{\partial^2 \varphi}{\partial z^2} + \varepsilon_{11}\left(\frac{\partial^2 \varphi}{\partial x^2} + \frac{\partial^2 \varphi}{\partial y^2}\right) = \frac{1}{\varepsilon_0}\frac{\partial P_3}{\partial z}, \\ \varphi(x,y,z=0) = V_e(x,y), \quad \varphi(x,y,z=h) = 0. \end{cases} \quad (1)$$

Here we introduced the dielectric permittivity of the background [50] or reference state [51] as $\varepsilon_{33}^b$ (typically $\varepsilon_{33}^b \leq 10$). $V_e(x,y)$ is the potential distribution at the sample surface; $\varepsilon_0$ is the universal dielectric constant; *h* is the sample thickness.

The electrostatic potential $\varphi(\mathbf{r})$ includes the effects of the probe field as well as the depolarization field created by the bound polarization charges of the counter wall at the domain apex. The perfect screening of the depolarization field [52] outside the sample is realized by the ambient screening charges.

In the effective point charge approximation, the potential distribution produced by the SPM probe on the surface of semi-infinite sample can be approximated as $V_e(x,y) \approx V d/\sqrt{x^2+y^2+d^2}$, where *V* is the applied bias, *d* is the effective charge-surface separation determined by the probe geometry [see Ref. [36, 53, 54] and Fig. 1(a)]. The potential is normalized assuming the condition of a perfect electrical contact with the surface, $V_e(0,0) \approx V$. In the case of a flattened tip represented by a disk of radius $R_0$ in contact with the sample surface, separation $d = 2R_0/\pi$ and is almost independent on the film depth and its dielectric permittivity [55].

In the framework of the LGD phenomenology, a stable or metastable polarization distribution inside the proper ferroelectric can be found as the solution of the stationary LGD equation:

$$\alpha P_3 + \beta P_3^3 + \delta P_3^5 - \xi \frac{\partial^2 P_3}{\partial z^2} - \eta \left( \frac{\partial^2 P_3}{\partial x^2} + \frac{\partial^2 P_3}{\partial y^2} \right) = -\frac{\partial \varphi}{\partial z} . \qquad (2a)$$

The gradient (or correlation) terms $\xi > 0$ and $\eta > 0$ (usually $\xi \sim \eta$), the expansion coefficient $\delta > 0$, while $\beta < 0$ ($\beta > 0$) for the first (second) order phase transitions. The coefficient $\alpha < 0$ in ferroelectric phase. Rigorously speaking, the coefficient $\alpha$ should be renormalized by the elastic stress (in e.g. thin films) [56, 57].

The boundary conditions for the polarization distribution are:

$$P_3(r \gg d, z < 0) \to -P_S, \qquad \frac{\partial P_3}{\partial z}(z = 0) = 0, \qquad \frac{\partial P_3}{\partial z}(z = h) = 0, \qquad (2b)$$

where $P_S$ is the initial spontaneous polarization value. The boundary condition $\partial P_3/\partial z = 0$ is called "natural" [50, 58] and corresponds to the case, when one could neglect the surface energy contribution and use $\lambda \to \infty$ in a more general condition $P_3 + \lambda(\partial P_3/\partial z) = 0$. In the case of the natural boundary conditions, a constant polarization value $P_3 = P_S$ satisfies Eq. (2b) at zero external bias, $V=0$. For the first order ferroelectric, the spontaneous polarization in the bulk is $P_S^2 = \left( \sqrt{\beta^2 - 4\alpha\delta} - \beta \right) / 2\delta$, while $P_S^2 = -\alpha/\beta$ for the second order ferroelectric [47].

The stationary solution of Eq.(2) is the extremum of the free energy with respect to polarization, $\delta G[P_3]/\delta P_3 = 0$, where

$$G[P_3] = \int_{-\infty}^{\infty} dx \int_{-\infty}^{\infty} dy \int_0^{\infty} dz \left( \frac{\alpha}{2} P_3^2 + \frac{\beta}{4} P_3^4 + \frac{\delta}{6} P_3^6 + \frac{\xi}{2} \left( \frac{\partial P_3}{\partial z} \right)^2 + \frac{\eta}{2} (\nabla_\perp P_3)^2 - P_3 \left( E_3^e + \frac{E_3^d}{2} \right) \right) \qquad (3)$$

Here $E_3^d$ is the depolarization field, $E_3^e$ is the external (probe) electric field.

**3. Polarization redistribution induced by small probe bias: subcritical nucleus**

To obtain the spatial distribution of the polarization at small positive biases, $V$, Eq. (2a) is linearized as $P_3(\mathbf{r}) = -P_S + p(\mathbf{r})$, where $p(\mathbf{r})$ is the induced polarization field due to materials response to a biased probe. The condition $p(\mathbf{r}) \to 0$ is valid far from the probe at an arbitrary applied bias. Here, we derive the solution within a perturbation approach.

Under the condition of a thick film, $h \gg d$, the approximate closed form expression for the linearized solution of Eq. (2a) is derived as (see Supplement for details):

$$p(\rho, z) \approx \frac{V}{(d+L_\perp)\alpha_S} \left( \frac{(d+z/\gamma)d^2}{\left(L_\perp(d+z/\gamma) + (d+z/\gamma)^2 + \rho^2\right)^{3/2}} + \frac{d^2(d^2+\rho^2) - 3d^4}{\gamma(d^2+\rho^2)^{5/2}} L_z \exp\left(-\frac{z}{L_z}\right) \right).$$

(4)

Here $\rho = \sqrt{x^2 + y^2}$ is the radial coordinate. The length $L_\perp = \sqrt{\eta/\alpha_S}$ defines the finite intrinsic width of the $180^\circ$-domain wall, where the renormalized coefficient $\alpha_S = \alpha + 3\beta P_S^2 + 5\delta P_S^4$. The correlation length $L_z = \sqrt{\varepsilon_0 \varepsilon_{33}^b \xi}$ is extremely small for typical values of $\xi \sim 10^{-8} \ldots 10^{-10}$ J m$^3$/C$^2$ in SI units. The effective dielectric anisotropy factor $\gamma = \sqrt{\gamma_b^2 + 1/(\varepsilon_{11}\varepsilon_0 \alpha_S)}$ and the "bare" dielectric anisotropy factor $\gamma_b = \sqrt{\varepsilon_{33}^b/\varepsilon_{11}}$ are introduced.

When deriving expression (4), we utilized the inequalities $2\varepsilon_0 \varepsilon_{33}^b |\alpha_S| \ll 1$, $\varepsilon_{33}^b \ll \varepsilon_{33}$, $L_\perp \le 0.5\ldots5$ nm and $L_z < 1$ Å, valid for typical ferroelectric material parameters and the background permittivity $\varepsilon_{33}^b \le 5$. Assuming the validity of additional inequalities $L_z \ll L_\perp \ll d$, the approximate solution of Eq.(1)-(2) was derived in the Supplement S.1 as:

$$P_3(\rho, z) \approx -P_S + \frac{V}{\alpha_S d} \frac{(d+z/\gamma)d^2}{\left((d+z/\gamma)^2 + \rho^2\right)^{3/2}}, \quad \text{at} \quad z \gg L_z, \quad (5a)$$

$$E_3(\rho, z) = -\frac{\partial \varphi(\rho, z)}{\partial z} = \frac{V(d+z/\gamma)d}{\gamma\left((d+z/\gamma)^2 + \rho^2\right)^{3/2}}. \quad (5b)$$

The profiles of the probe-induced polarization redistribution calculated within the LGD approach from the analytical expression in Eq.(4) are shown in Figs. 2. It is clear that the aspect ratio of the domain nucleus is close to the dielectric anisotropy factor γ. Moreover, the polarization distribution inside the *subcritical* domain nucleus is very smooth or "soft" and no sharp changes (and thus strong depolarization field) appear.

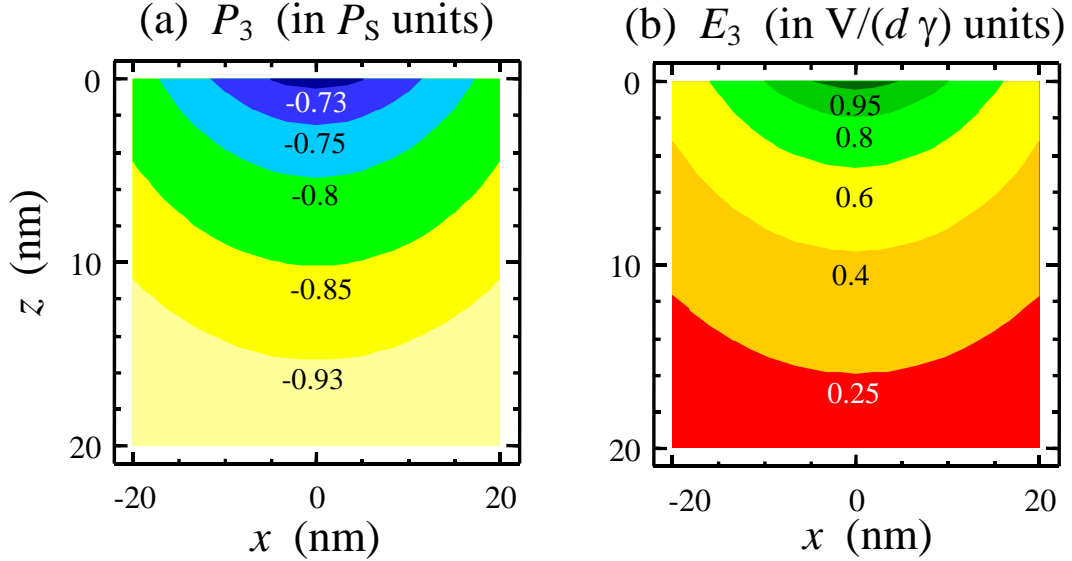

**FIG. 2**. Probe-induced polarization $P_3(x,z)$ (a) and external field (b) distribution in LiNbO$_3$ calculated from Eq.(4) for applied bias 20 V before the domain formation. Figures near the contours are polarization values in $P_S$ units. Material parameters: $\varepsilon_{11} = 84$, $\varepsilon_{33} = 30$, $\alpha = -1.95 \cdot 10^9$ m/F, $\beta = 3.61 \cdot 10^9$ m$^5$/(C$^2$F), $P_S = 0.73$ C/m$^2$ [59], gradient coefficients $\eta = \xi = 10^{-9}$ SI units; effective distance $d = 25$ nm, $\varepsilon_{33}^b \leq 5$.

The linear approximation for the polarization distribution given by Eq.(4) is quantitatively valid until $|p| \ll P_s$ or, alternatively, $V/d \ll \alpha_s P_s$, i.e. at biases $V$ much smaller than the coercive bias, at which polarization reversal is absent. This means that the probe induced domain formation cannot be considered quantitatively within the linearized LGD-equation. Below we take into account the ferroelectric material nonlinearity.

## 4. Probe-induced domain formation and ferroelectric breakdown through the film

### *4.1. Analytical solution*

The analysis of the domain switching beyond the linear models requires the analytical description of the depolarization field produced by the counter domain wall. Since for most ferroelectrics the tip size is larger then correlation length, $L_\perp \ll d$, this approximation is used hereinafter. Applying the direct variational method [49] to the simplified linearized solution Eq. (5a), we obtained that in the actual region $z \gg L_z$, the gradient effects lead to the unessential renormalization of expansion coefficient $\alpha$ as $\alpha \to \alpha_R = \alpha(1 + (1+\gamma^{-2})L_\perp^2/d^2)$. The approximation is rather rigorous outside the domain wall region. Hence, coupled Eqs.(1)-(2) can be rewritten as

$$\begin{cases} \gamma_P^2(P_3)\frac{\partial^2 \varphi}{\partial z^2} + \frac{\partial^2 \varphi}{\partial x^2} + \frac{\partial^2 \varphi}{\partial y^2} = 0, \\ \alpha_R P_3 + \beta P_3^3 + \delta P_3^5 = -\frac{\partial \varphi}{\partial z}, \\ \varphi(x,y,z=0) = \frac{Vd}{\sqrt{x^2+y^2+d^2}}, \quad \varphi(x,y,z=h)=0. \end{cases} \quad (6)$$

Polarization-dependent anisotropy factor $\gamma_P[P_3]$ in Eqs. (6) is formally introduced as:

$$\gamma_P(P_3) = \sqrt{\frac{\varepsilon_{33}^b}{\varepsilon_{11}} + \frac{1}{\varepsilon_{11}\varepsilon_0(\alpha_R + 3\beta P_3^2 + 5\delta P_3^4)}}. \quad (7)$$

Note that the dielectric susceptibility $\chi \sim (\alpha_R + 3\beta P_3^2 + 5\delta P_3^4)^{-1}$ is positive for thermodynamically stable states. In order to obtain analytical results, we regard $\gamma_P \approx \gamma$ that is equivalent to neglecting the electric field dependence of the linear permittivity.

The spatial distribution of the z-component of the electric field can be represented as

$$E_3(\rho,z) = E_P(\rho,z) + E_W(\rho,z). \quad (8)$$

Where $E_P(\rho,z)$, is the probe field inside the sample and $E_W(\rho,z)$ is the depolarization field the due to the charged domain wall, which is absent in the linearized solution (5b) for the "soft" subcritical nucleus. The term $E_P(\rho,z)$ is:

$$E_P(\rho,z) = Vd \int_0^\infty dk J_0(k\rho) \exp(-kd) \frac{\cosh(k(h-z)/\gamma)}{\sinh(kh/\gamma)} \frac{k}{\gamma} \quad (9a)$$

The integral in Eq. (9a) can be expanded in the image charge series. For very thick ($h \gg \gamma d$) or ultra-thin ($h \ll \gamma d$) films, the series was reduced to the first term to obtain the approximate expressions:

$$E_P(\rho,z) \approx \begin{cases} \dfrac{V}{\gamma} \dfrac{(d+z/\gamma)d}{\left((d+z/\gamma)^2+\rho^2\right)^{3/2}}, & h \gg \gamma d, \\[2ex] \dfrac{Vd}{2h}\left(\dfrac{1}{\sqrt{(d-(h-z)/\gamma)^2+\rho^2}} + \dfrac{1}{\sqrt{(d+(h-z)/\gamma)^2+\rho^2}}\right), & h \ll \gamma d. \end{cases} \quad (9b)$$

When the domain nucleus appears, the domain wall containing the uncompensated bound electric charge with the total surface density of $\sigma_b(\mathbf{r}) = 2P_S n_z(z)$ [60] and screening charge with density $\sigma_S(\mathbf{r}) = \sigma \cdot n_z(z)$, originated from the band bending effect (if the latter is strong enough), produce the additional depolarization field $E_W(\rho,z)$.

For the case $l \ll h$, the value of $E_W(\rho,z)$ was analytically calculated in the approximation of the semi-ellipsoidal domain with radius *r*, length *l* and the *finite intrinsic width* of the curved domain wall estimated as $L_W(z) \approx L_\perp \sqrt{1+(z/r)^2-(z/l)^2}$ (Suppl. 2). For the case:

$$E_W = \frac{\varphi_L(\rho,z) - \varphi_L(\rho, z+L_W)}{L_W(z)} \approx \begin{cases} -\Delta E \cdot n_D(a), & \rho \leq r - L_\perp, \ z=0, \\ \Delta E\left(f_D\left(a, \dfrac{L_\perp l}{r}\right) - n_D(a)\right), & \rho = 0, \ l-L_\perp < z < l, \end{cases} \quad (10a)$$

Where $\Delta E = (2P_S+\sigma)/(\varepsilon_0 \varepsilon_{11} \gamma^2)$ is the field amplitude, $a = r\gamma/l$ is the domain aspect ratio, $n_D(a) = a^2(1-a^2)^{-3/2}\left(\text{arctanh}\sqrt{1-a^2} - \sqrt{1-a^2}\right)$ is the depolarization factor [61]. Approximately, $n_D(a) \sim a^2/(1+a^2)$. The function $f_D(a, L_W)$ is given by expression

$$f_D(a, L_W) = \left(1+\frac{l}{L_W}\right)\left(n_D(a) - \frac{a^2}{(1-a^2)^{3/2}}\left(\text{arctanh}\frac{\sqrt{1-a^2}}{1+L_W/l} - \frac{\sqrt{1-a^2}}{1+L_W/l}\right)\right). \quad (10b)$$

The expression for the Landauer depolarization field potential $\varphi_L(\rho,z)$ is well-known [39] and can be written as:

$$\varphi_L(\rho, z) = \Delta E \frac{a^2 z}{(1-a^2)^{3/2}} \left( \operatorname{arctanh}\left( \sqrt{\frac{1-a^2}{1+\zeta\gamma^2/l^2}} \right) - \sqrt{\frac{1-a^2}{1+\zeta\gamma^2/l^2}} \right),$$

$$\zeta = \begin{cases} 0, & \dfrac{\rho^2}{r^2} + \dfrac{z^2}{l^2} < 1 \\ \dfrac{1}{2}\left( \sqrt{\left(r^2 - \rho^2 + \dfrac{z^2 - l^2}{\gamma^2}\right)^2 + 4\rho^2 \dfrac{z^2}{\gamma^2}} + \rho^2 - r^2 + \dfrac{z^2 - l^2}{\gamma^2} \right), & \dfrac{\rho^2}{r^2} + \dfrac{z^2}{l^2} \geq 1 \end{cases} \quad (11)$$

It is very important for further logic, that the field $E_W(\rho, z)$ (given by Eqs.(10) for the finite width $L_\perp > 0$) differs from the Landauer depolarization field $E_L(\rho, z) = -\partial \varphi_L / \partial z$ (corresponding to the case of infinitely-thin domain walls with $L_\perp = 0$). The Landauer field is homogeneous inside the semi-ellipsoidal domain and vanishes as $(r/l)^2$ at $r/l \to 0$. However, outside the domain tip it changes the sign (allowing for the surface bound charge) and so it acts as the polarizing field that can exceed the intrinsic coercive field $E_c$, for the second order ferroelectrics $E_c = P_S / (3\sqrt{3}\varepsilon_0 \varepsilon_{11} \gamma^2)$ [see filled regions in Figs.3 (a,b)]. Note that at the domain face z=0 the field $E_L(\rho, z)$ is continuous, while at the domain tip the jump has appeared:

$$E_L(r-0,0) = E_L(r+0,0) = -n_D(a)\Delta E,$$

$$E_L(0, z = l - 0) = -n_D(a)\Delta E, \quad E_L(0, z = l + 0) = (1 - n_D(a))\Delta E. \quad (12)$$

The jump of depolarization field $E_L(0, l)$ for the case of an infinitely thin counter domain wall is illustrated in Fig. 3(c).

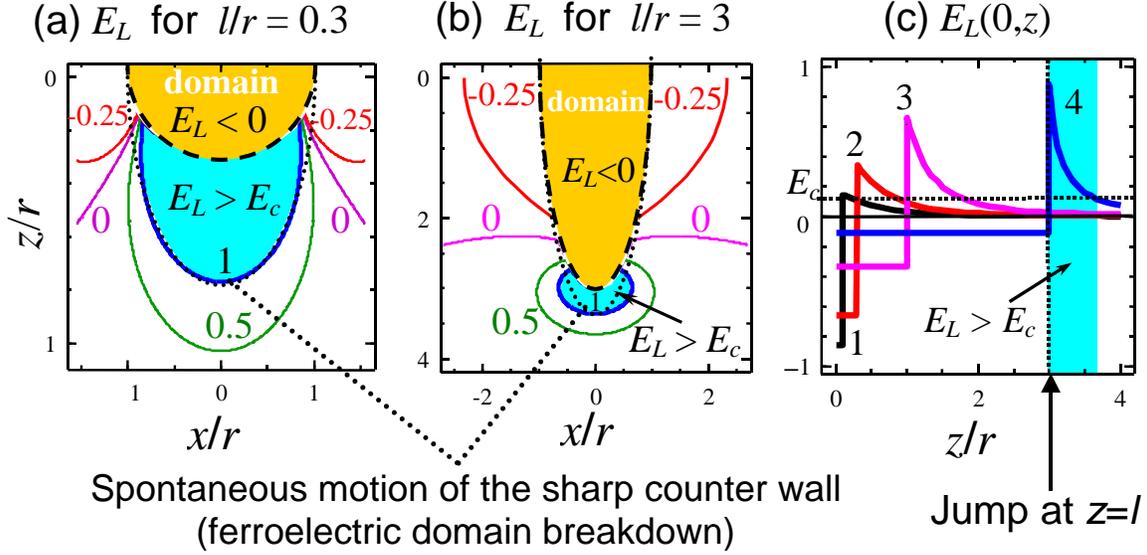

**FIG.3**. (a-c) Mechanism of domain breakdown in the case of infinitely-thin counter domain wall. (a,b) Constant lines of the Landauer depolarization field ratio $E_L(\rho,z)/E_c$ for semi-ellipsoidal domain with radius $r$ and length $l$ in LiNbO$_3$ with $\sigma = 0$. Dashed contour is the initial domain boundary. Filled areas indicate the region where depolarization field $E_L$ is more than coercive field $E_c$. Dotted contour schematically shows the new domain boundary originated from the polarizing effect of the counter domain wall. (c) Depth distribution $E_L(0,z)/\Delta E$ for different aspect ratios $l/r = 0.1,\ 0.3,\ 1,\ 3$ (curves 1-4). Horizontal line corresponds to coercive field $E_c$.

The strong *positive* depolarization field ($E_L > E_c$) in front of the infinitely-thin charged domain wall causes the spontaneous increase of the domain length leading to the domain wall breakdown into the depth of the sample (compare with the spike-like domain appearance and domain breakdown calculated within energetic LM-approach). Complementary to the LM approach evolved for infinitely thin domain walls, our approach provides the solution of the paradox: the domain vertical growth should be accompanied by the increase of the width of the charged domain wall. Actually, the wall width near the domain tip $z=l$ increases with domain length $l$ increase as $L_W(l) \cong L_\perp\, l/r$, but the width of the 180°-domain wall, $L_W(0) = L_\perp$ located at the domain face $z=0$ remains constant in accordance with expression for domain wall width $L_W(z) \approx L_\perp\sqrt{1+(z/r)^2-(z/l)^2}$ valid for quasi-ellipsoidal domain shape. The width increase smears the jump of the depolarization field at the domain tip, and the domain wall broadening and propagation is finished once the field in front of the wall becomes smaller than the coercive field. Note, that mathematically Eqs. (10) is similar to the *averaging of* depolarization field over the domain wall as proposed and argued by Drugard and Landauer for the flat domain wall [62].

Using Eqs.(10), the spatial distribution of the polarization can be found as the solution of the nonlinear algebraic equation

$$\alpha_R P_3(\rho,z) + \beta P_3^3(\rho,z) + \delta P_3^5(\rho,z) = E_3(\rho,z). \tag{13}$$

We emphasize that the effective field $E_3$ is the sum of the probe and depolarization fields. The left-hand-side of Eq. (13) describes the conventional ferroelectric hysteresis. Thus, under the *absence of the pinning field*, a thermodynamically stable domain wall boundary $\rho(z)$ can be determined from the Eq.(13) as the coercive point, i.e. under the condition $\left(\alpha_R + 3\beta P_3^2(\rho,z) + 5\delta P_3^4(\rho,z)\right) = 0$ valid at coercive field: $E_3(\rho,z) = E_c$.

The intrinsic coercive field $E_c$ is well-known [63] as:

$$E_c = \begin{cases} \dfrac{2}{3\sqrt{3}}\sqrt{-\dfrac{\alpha_R^3}{\beta}}, & \text{for the second order ferroelectrics,} \\ \dfrac{2}{5}\left(2\beta + \sqrt{9\beta^2 - 20\alpha_R\delta}\right)\left(\dfrac{2\alpha_R}{-3\beta - \sqrt{9\beta^2 - 20\alpha_R\delta}}\right)^{3/2} & \text{for the first order.} \end{cases} \tag{14}$$

Note that this analysis essentially reproduces early arguments of Kolosov [64], stating that the domain size in a PFM experiment corresponds to the region in which tip-induced field exceeds coercive field. Here, we obtain a similar result; however, the field is now intrinsic (rather then macroscopic) coercive field renormalized by the depolarization field of the nascent domain.

### *4.2. Vertical growth of the domain in thick films*

The bias dependence of the domain radius $r(V)$ at the sample surface should be determined from the equation $E_3(\rho,z) = E_c$ at $z = 0$, while the domain length $l(V)$ is determined at $\rho = 0$. For film with thickness $h \gg d$ and domain length $l \ll h$ we derived coupled equations for the radius *r* and length *l* bias dependences:

$$\begin{cases} E_P(r,0) - n_D\left(\dfrac{\gamma r}{l}\right)\Delta E = E_c, \\ E_P(0,l) + \Delta E\left(f_D\left(\dfrac{\gamma r}{l}, \dfrac{L_\perp l}{r}\right) - n_D\left(\dfrac{\gamma r}{l}\right)\right) = E_c. \end{cases} \quad (15)$$

Here the approximate analytical expressions for the probe field $E_P$ is given by Eq. (9), the factor $0 < f_D < 1$ is given by Eq.(10b) and $\Delta E = (2P_S + \sigma)/(\varepsilon_0 \varepsilon_{11} \gamma^2)$. As anticipated, the domain breakdown through the sample depth ($l \to \infty$) appears under the condition $f_D \Delta E > E_c$ which is true for a negligible intrinsic width $L_\perp \to 0$.

When the domain approaches the bottom electrode (oppositely to the above-considered case $l \ll h$) we put $l = h$ and $f_D = 0$ in Eqs.(15), and thus obtained rough estimations for the corresponding domain radius and critical bias that initiates domain intergrowth through the sample depth:

$$r_{int}(h) = \begin{cases} d\sqrt{(1+(h/\gamma d))^{4/3} - 1} \sim h^{2/3}, & h > \gamma d \\ (h/\gamma)\sqrt{2 - 7(h/\gamma d)^2} \sim h, & h \ll \gamma d \end{cases}, \quad (16a)$$

$$V_{int}(h) \approx \begin{cases} \gamma d\left(1 + \dfrac{h}{\gamma d}\right)^2 \left(E_c + n_D\left(\dfrac{\gamma r_{int}}{h}\right)\Delta E\right) \sim h^{4/3}, & h > \gamma d, \\ h\left(E_c + n_D\left(\dfrac{\gamma r_{int}}{h}\right)\Delta E\right) \sim h, & h \ll \gamma d. \end{cases} \quad (16b)$$

Note, that expressions (16) derived for the case of the electric excitation by the localized probe field with characteristic scale $d$ differs from the semi-empirical Kay-Dunn law, which stated that $r \sim h^{2/3}$ and coercive field $E_{cr} \sim h^{-2/3}$ for homogeneous external field.

For films with thickness $h \gg \gamma d$, the bias dependences of the domain length $l(V)$ and radius $r(V)$ calculated from Eqs. (15) are shown in Fig. 4 for LiNbO$_3$ and Fig. 5 for typical ferroelectric materials including LiTaO$_3$, PbTiO$_3$ and PbZr$_{40}$Ti$_{60}$O$_3$ in three limiting cases.

(i) Perfect screening of domain wall depolarization field by free charges $\sigma = -2P_S$. For this case there are no resulting charge at the wall and no depolarization field, (see **dashed** curves in Figs.4a,b and 5).

(ii) No motion of the charged domain wall by depolarization field ($f_D = 0$) and no screening charges ($\sigma = 0$). This case has unclear physical interpretation and shown by **dotted** curves in Figs.4a,b and 5 for comparison only.

(iii) The motion of the charged domain wall by the maximal depolarization field is considered ($\sigma = 0$, $f_D > 0$). The situation is typical in the absence of screening or very slow screening (see **solid** curves in Figs.4a,b and 5).

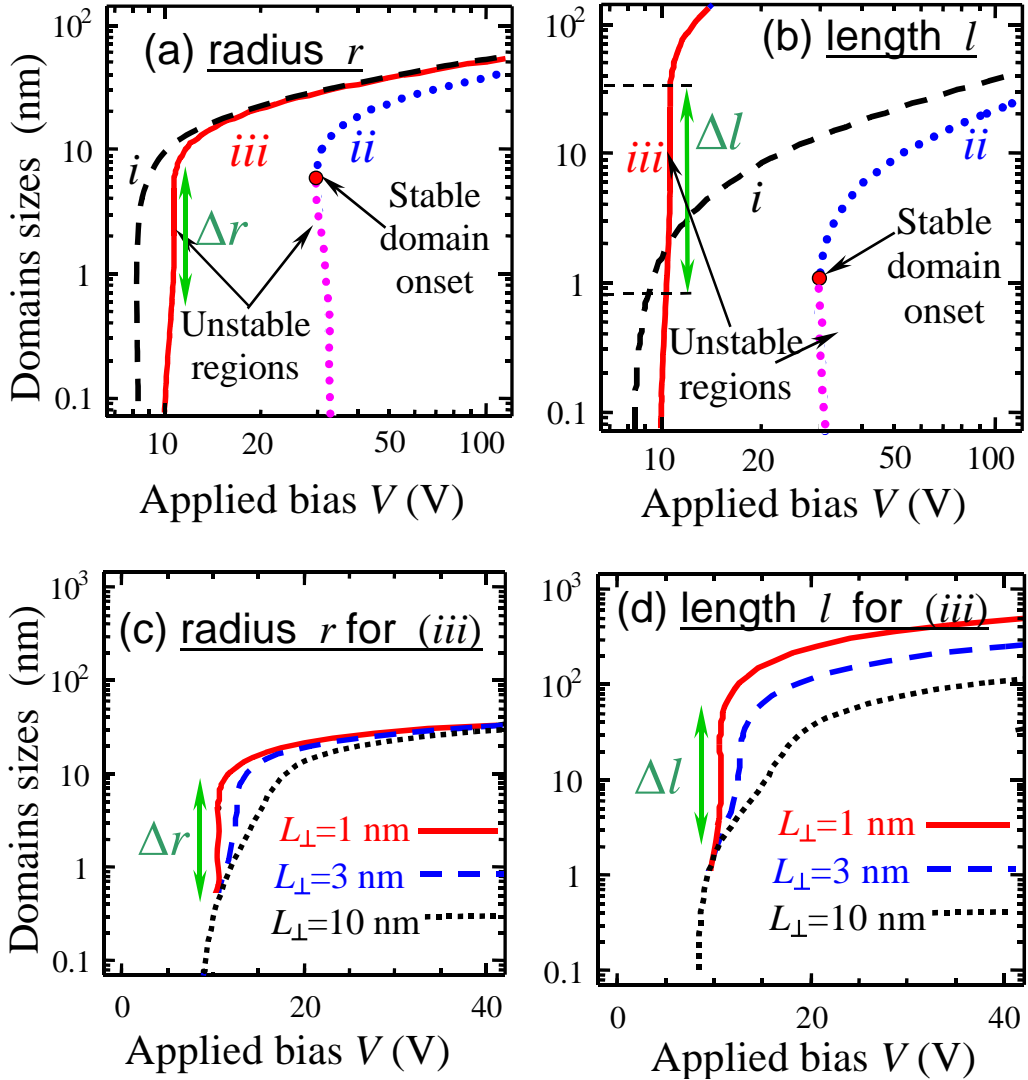

**FIG. 4**. Domain radius $r(V)$ (a,c) and length $l(V)$ (b,d) bias dependence calculated within LGD-approach for LiNbO$_3$ (with $\varepsilon_{11} = 84$, $\varepsilon_{33} = 30$, $\alpha = -1.95 \cdot 10^9$ m/F, $\beta = 3.61 \cdot 10^9$ m$^5$/(C$^2$F), $P_S = 0.73$ C/m$^2$). Effective distance $d = 25$ nm, $\varepsilon_{33}^b \leq 5$, sample thickness $h \to \infty$. (a,b) Solid curves

are calculated from Eqs.(15) for $\sigma = 0$, $f_D > 0$ (case iii), $L_\perp = 1$ nm; dashed curves correspond to $\sigma \to -2P_S$ (case i); dotted curves correspond to $\sigma = 0$ and $f_D = 0$ (case ii). (c,d) Solid, dashed and dotted curves correspond to the case (iii) and $L_\perp = 1, 3, 10$ nm.

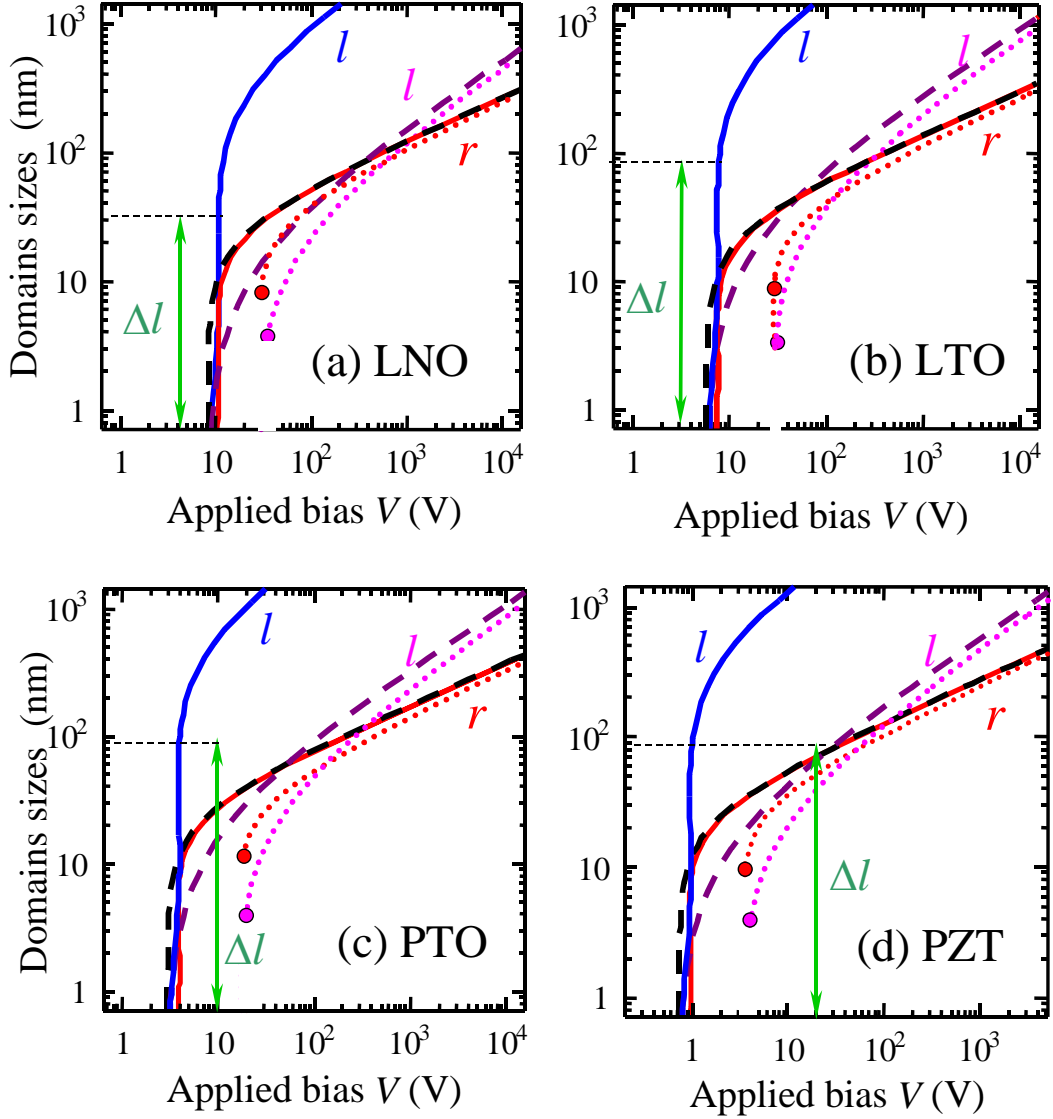

**FIG. 5**. Domain length $l(V)$ and radius $r(V)$ bias dependence calculated within LGD-approach for typical ferroelectric materials: LNO (LiNbO$_3$ with $\varepsilon_{11} = 84$, $\varepsilon_{33} = 30$, $\alpha = -1.95 \cdot 10^9$ m/F, $\beta = 3.61 \cdot 10^9$ m$^5$/(C$^2$F), $P_S = 0.73$ C/m$^2$); LTO (LiTaO$_3$ with $\varepsilon_{11} = 54$, $\varepsilon_{33} = 44$, $\alpha = -1.31 \cdot 10^9$ m/F, $\beta = 5.04 \cdot 10^9$ m$^5$/(C$^2$F), $P_S = 0.51$ C/m$^2$); PTO (PbTiO$_3$ with $\varepsilon_{11} = 124$, $\varepsilon_{33} = 67$, $\alpha = -3.42 \cdot 10^8$ m/F, $\beta = -2.90 \cdot 10^8$ m$^5$/(C$^2$F), $\delta = 1.56 \cdot 10^9$ m$^5$/(C$^2$F), $P_S = 0.75$ C/m$^2$); PZT (PbZr$_{40}$Ti$_{60}$O$_3$ with $\varepsilon_{11} = 497$,

$\varepsilon_{33} = 197$, $\alpha = -1.66 \cdot 10^8$ m/F, $\beta = 1.44 \cdot 10^8$ m$^5$/(C$^2$F), $\delta = 1.14 \cdot 10^9$ m$^5$/(C$^2$F), $P_S = 0.57$ C/m$^2$ [65]). Effective distance $d = 25$ nm, $\varepsilon_{33}^b \leq 5$, $L_\perp = 1$ nm, sample thickness $h \to \infty$. Solid curves are calculated from Eqs.(14) for $\sigma = 0$, $f_D > 0$ (case iii), dashed curves correspond to $\sigma \to -2P_S$ (case i); dotted curves correspond to $\sigma = 0$ and $f_D = 0$ (case ii).

The calculated coercive biases $V_c \sim 1\text{-}10$ V of domain reversal are in the same range as available experimental results [28-31, 66, 67], but further comparison is hindered by the lack of knowledge on the exact tip geometry. At biases $V < V_c$ the domain nucleation is absent in a real time scale. Under the perfect screening of domain wall depolarization field by free charges, the domain formation at biases $V \geq V_c$ is activationless, since domain appears with zero sizes $r(V_c) = l(V_c) = 0$ (see dashed curves in Fig.4(a)). In contrast, when the motion of the charged domain wall by depolarization field is absent, activation barrier appears, since unstable regions appeared at the domain onset (see dotted curves at Fig.4(a)).

Note, that the behavior of the curves at sizes less that 0.8 nm shown in Fig. 4 should be ignored, since for the sizes less 2-3 lattice constants, the continuous LGD-approach is not valid. However, the jumps of the domain radius $\Delta r$ and length $\Delta l$ up to sizes more that tens nm should be interpreted as the first-order nucleation (see solid curves onset in Figs. 4-5). Figs.4 (c-d) could be interpreted as the continuous crossover between the first-order and the second-order domain nucleation appeared under $L_\perp$ increase in the strongly inhomogeneous probe field. Independently on the phase transition order, the activation barrier disappears at coercive bias $V_c$.

The approximate expressions for the domain radius $r$, length $l$ bias dependences and shape $\rho(z)$ derived from Eq.(15) are summarized in the Tab.1 for the cases (i)-(iii). Note, that the coercive bias $V_c$ of domain formation is proportional to the intrinsic coercive field $E_c$ given by Eq.(14).

**Table 1.**

| Domain | Intrinsic model of domain formation for thick films ($h \gg \gamma d$) |
|---|---|

| charac-teristics | Case (i): $\sigma = -2P_S$ (complete screening) | Case (ii): $\sigma > -2P_S$, $f_D = 0$ | Case (iii): $\sigma > -2P_S$, $f_D > 0$ (slow screening) |
|---|---|---|---|
| Coercive bias $V_c$ | $V_c = \gamma\, d \cdot E_c$ | $V_c = \gamma\, d(E_c + \Delta E)$, $\Delta E = (2P_S + \sigma)/(\varepsilon_0\varepsilon_{11}\gamma^2)$ | $\gamma\, d\, E_c < V_c < \gamma\, d(E_c + \Delta E)$, $\Delta E = (2P_S + \sigma)/(\varepsilon_0\varepsilon_{11}\gamma^2)$ |
| Domain onset at $V \to V_c$ | Onset is activationless, since oblate domain appears with zero sizes $r(V_c) = 0$ and $l \sim r^2$. | Activation barrier exists, since prolate stable domain appears with nonzero sizes $l(V_c) > r(V_c) > 0$. | Nucleus is prolate, $r(V_c) = 0$. Spike-like stable domain ($r/l \ll 1$) appears after the almost first-order transition (see vertical parts of $l$-curves) |
| Sizes $r$ and $l$ vs. bias | $r(V) = d\sqrt{(V/V_c)^{2/3} - 1}$, $l(V) = \gamma \cdot d\left(\sqrt{V/V_c} - 1\right)$, $r^3/l^2 \approx \gamma^{-2}$ at $V/V_c \gg 1$ | $l = \gamma\, d\left((1 + r^2/d^2)^{3/4} - 1\right)$, $r(V \gg V_c) \sim V^{1/3}$, $l(V \gg V_c) \sim V^{1/2}$ | $r(V) \approx d\sqrt{(V/V_c)^{2/3} - 1}$, length $l \gg r$ is determined by $f_D$ value |
| Shape at $V > V_c$ | Equation for domain wall boundary: $\rho(z) = \begin{pmatrix} (V/V_c)^{2/3} d^{4/3} \times \\ \times (d + z/\gamma)^{2/3} \\ -(d + z/\gamma)^2 \end{pmatrix}^{1/2}$ | Domain is prolate. At high voltages $V/V_c \gg 1$ the invariant $r^3/l^2 \approx \gamma^{-2}$ exists (compare with invariant $r^3/l^2 \approx$ const obtained by Molotskii [44]). | Domain is strongly prolate. Domain breakdown through the sample depth ($l \to \infty$) appears under the condition $L_\perp \to 0$. Domain length decreases with $L_\perp$ increase |

As it follows from Tab.1, under the absence of the domain wall motion by the depolarization field, the domain length depends on bias as $l(V) \sim V^{1/2}$ at high voltage, while the domain radius $r(V) \sim V^{1/3}$ increases more slowly than in LM approach with $l(V) \sim V$ and $r(V) \sim V^{2/3}$. If the strong positive depolarization field moves the charged domain wall, we still obtained that $r(V) \sim V^{1/3}$, but the domain length rapidly increases.

### *4.3. Lateral growth of the domain in the film*

Finally, we consider the lateral growth of cylindrical domain appeared after the domain breakdown in thin ferroelectric films. Under the condition of the domain intergrowth through the film depth, the charged domain wall disappears (all walls are 180-degree) and so one should put $E_W = 0$ in Eq.(8) and (13). At finite film thickness, electric field (9) at the sample surface acquires the form:

$$E_P(\rho, z=0) = Vd\left(\frac{d}{\gamma(d^2+\rho^2)^{3/2}} + \sum_{n=1}^{\infty}\frac{2(d+2hn/\gamma)}{\gamma\left((d+2hn/\gamma)^2+\rho^2\right)^{3/2}}\right)$$

$$\approx \begin{cases} \dfrac{Vd}{\gamma(d^2+\rho^2)^{1/2}}\left(\dfrac{d}{(d^2+\rho^2)}+\dfrac{\gamma}{h}\right), & h \ll \gamma d \\ \dfrac{Vd}{\gamma(d^2+\rho^2)^{3/2}} + \dfrac{2V(d+2h/\gamma)}{\gamma\left((d+2h/\gamma)^2+\rho^2\right)^{3/2}}, & h \gg \gamma d \end{cases} \quad (17)$$

Using Eqs.(17) and (13), the domain radius dependence vs. bias and film thickness should be calculated from the equation $E_P(r,0) = E_c$. The approximate analytical expressions are

$$r \approx \begin{cases} d\sqrt{(V/E_c h)^2 - 1}, & h \ll \gamma d, \\ d\sqrt{(V/\gamma E_c d)^{2/3} - 1}, & h \gg \gamma d, \end{cases} \quad (18)$$

Note, that the dependences (18) are valid for domain lateral growth caused by strongly inhomogeneous probe electric field in ferroelectric film.

Bias dependences $r(V)$ are shown in Fig. 6. Note that the domain radius and the coercive voltage decrease with the film thickness. Obtained numerical values are in a reasonable agreement with Cho et al data [68, 69] for thin LTO films.

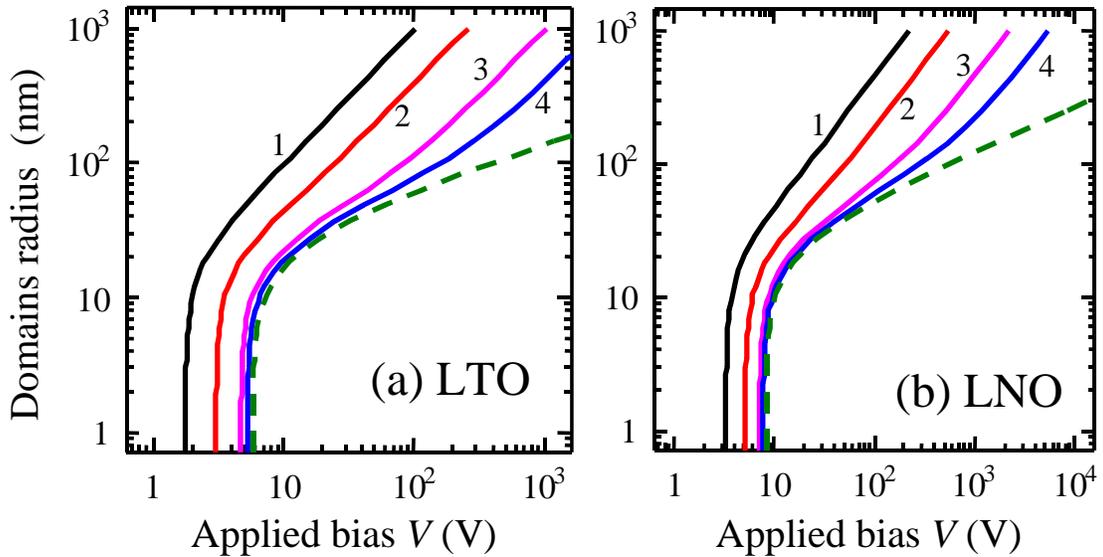

**FIG. 6**. Domain radius $r(V)$ bias dependence calculated within LGD-theory from for typical ferroelectric materials: LTO (a); LNO (b); materials parameters are listed in caption to Fig.4; Effective

distance $d = 25$ nm, $\varepsilon_{33}^b \leq 5$. Solid curves 1, 2, 3, 4 are calculated for different film thickness $h$: 10, 25, 100, 250 nm; dashed curves correspond to the dependence $r(V) = d\sqrt{(V/V_c)^{2/3} - 1}$ valid in semi-infinite sample after the domain breakdown.

The polarization distribution at the sample surface should be determined from the equation $\alpha_R P_3(\rho,0) + \beta P_3^3(\rho,0) + \delta P_3^5(\rho,0) = E_P(\rho,0)$. The bias dependence of the maximal polarization $P_3(0,0)$ in the center of cylindrical domain is shown in Fig. 7 for typical ferroelectric materials and film thickness $h = 250$ nm. Solid curves in Figs. 7(a,b) correspond to the stable dependence $P_3(V)$ and dashed curves are unstable states.

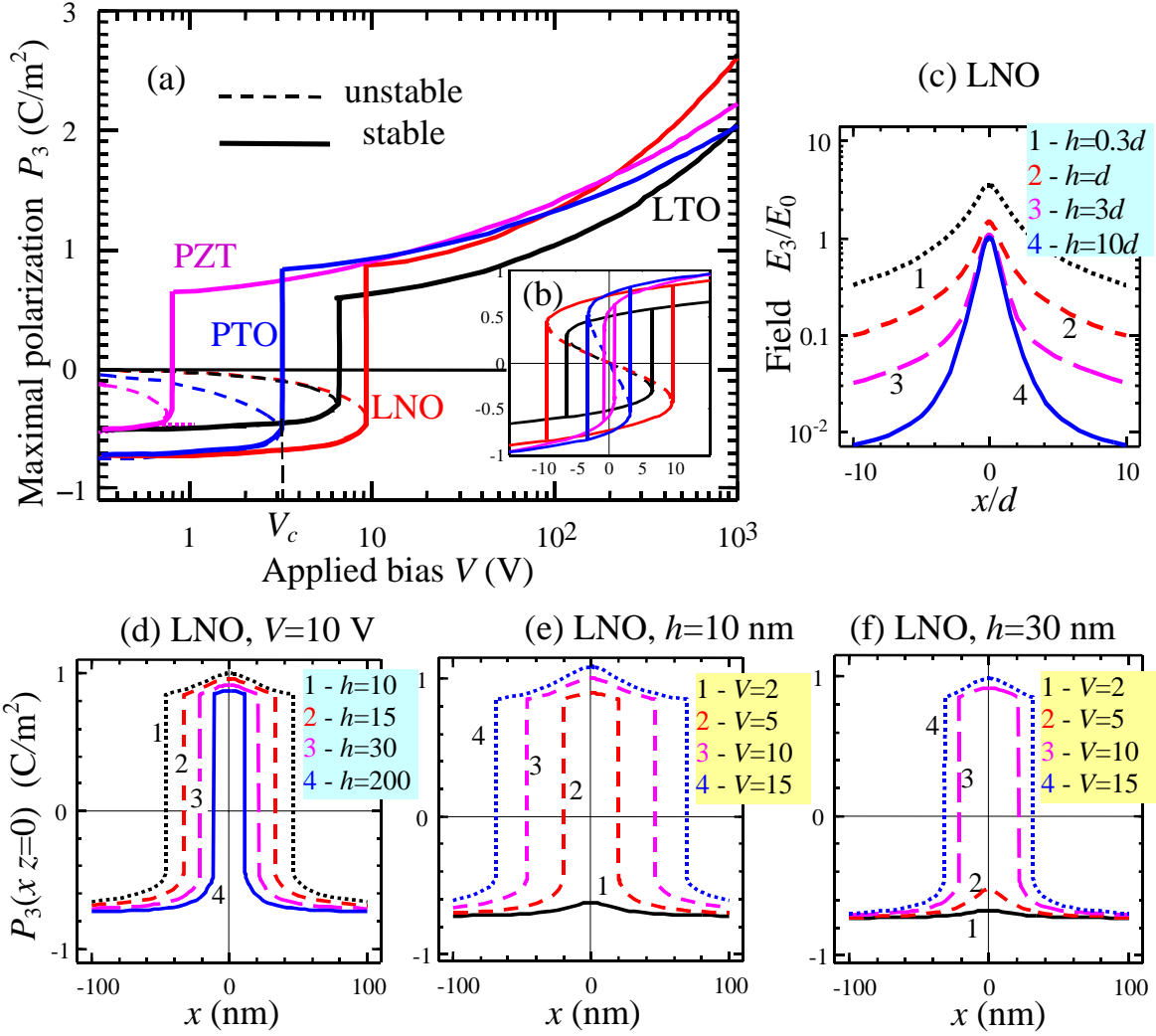

**FIG. 7**. (a) Polarization $P_3$ (0,0) in the center of cylindrical domain via increasing bias $V$ in linear-log scale for typical ferroelectric materials: PZT, PTO, LTO and LNO. (b) Ferroelectric hysteresis in linear scale. Material parameters and probe characteristics are the same as in Fig.4, $\sigma = 0$ and $l = h = 250$ nm. (c) Electric field $E_3(x,0)$ normalized on $E_0 = V/d$, (d-f) polarization $P_3(x,0)$ lateral distribution at the film surface z=0 for LiNbO$_3$. (c) Curves 1, 2, 3, 4 correspond to $h/d$=0.3, 1, 3,10. (d) Curves 1, 2, 3, 4 correspond to bias 10 V and different film thickness $h$=10, 15, 30, 200 nm. (e, f) Curves 1, 2, 3, 4 correspond to applied bias 2, 5, 10, 15 V and film thickness $h$=10 nm (e), 30 nm (f).

Note that the high PFM response contrast is possible under the condition $P_3 > 2P_S$ obtained within the LGD-approach with the increase of the applied bias. This opens a pathway for high-density data storage in ultra-thin ferroelectric layer.

Spatial distributions of the electric field (a) and (b-d) polarization $P_3(x,0)$ calculated at the sample surface are shown in Figs. 7c-f. In contrast to the smooth profile of the electric field and the "soft" polarization distribution inside the subcritical domain nucleus, the polarization distribution inside the stable cylindrical domain is rectangular-like or "hard" with rather sharp domain wall (compare curve 1 in Fig.7(e) and curves 1,2 in Fig.7(f) with curves 3-4).

## 6. Discussion

The remarkable aspect of the above analysis is that the domain radius $r$ calculated from Eqs. (15) is always finite at finite intrinsic domain wall width $L_\perp \neq 0$. This reflects the fact that spontaneous polarization re-orientation takes place inside the localized spatial region, where the resulting electric field absolute value is more that thermodynamic coercive field, i.e. $|E_3| > E_c$, while the hysteresis phenomenon appeared in the range $|E_3| < E_c$ as anticipated within LGD approach considering nonlinear correlation effects. The domain breakdown through the sample depth appears for infinitely thin domain walls ($L_\perp \to 0$), i.e. under the absence of domain wall correlation energy ($\xi, \eta \to 0$). The microscopic origin of the domain tip elongation in the region where the probe electric field is much smaller than the intrinsic coercive field is the positive depolarization field appearing in front of the moving charged domain wall. Note, that the *activationless* hysteresis phenomenon [e.g. shown in Fig.7 (b)] calculated within LGD approach corresponds to the *metastable state* [6], in contrast to activation mechanism of the stable domain formation calculated within energetic LM energetic approach. Thus, obtained results are complementary to the energetic approach.

As noted in the Introduction, within rigid LM-approach domain walls are regarded infinitely thin and polarization absolute value is constant: $-P_S$ outside and $+P_S$ inside the domain (if any). Semi-ellipsoidal domain radius $r$ and length $l$ are calculated from the free energy excess consisting of the interaction energy, the domain wall surface energy $\psi_S$ and the depolarization field energy (see S.3 and Refs. [32], [36], [44]). Nonlinear correlation energy contribution is absent within the rigid approximation. Within the LM-approach, the depolarization field energy vanishes as $1/l$, while the interaction energy is maximal at $l\to\infty$, the condition of negligible surface energy leads to the domain breakdown $l\to\infty$ and the subsequent macroscopic region re-polarization even at infinitely small bias (if only $VP_S > 0$), while the hysteresis phenomena or threshold bias (saddle point) are absent [32]. Under finite domain wall energy, the critical bias $V_{cr}$ and energetic barrier $E_a$ of stable domain formation exist. Activation (or nucleation) bias $V_a$ is determined from the condition $E_a(V_a) = n\, k_B T$, where the numerical factor $n = 1\ldots25$. Usually $V_a \gg V_{cr}$ for thick films [36].

In the Fig. 8 we compare the main features of the probe-induced domain formation calculated within intrinsic LGD-approach and energetic LM-approach. For consistency between the approaches we used the Zhirnov expression for the domain wall surface energy

$$\psi_S = \sqrt{\left(1 + \frac{2\left((Q_{11}^2 + Q_{12}^2)s_{11} - 2Q_{11}Q_{12}s_{12}\right)}{\beta(s_{11}^2 - s_{12}^2)}\right)\eta}\, \frac{(-2\alpha)^{3/2}}{3\beta}, \qquad (19)$$

where $Q_{ij}$ are electrostriction tensor, $s_{ij}$ are elastic compliances [70].

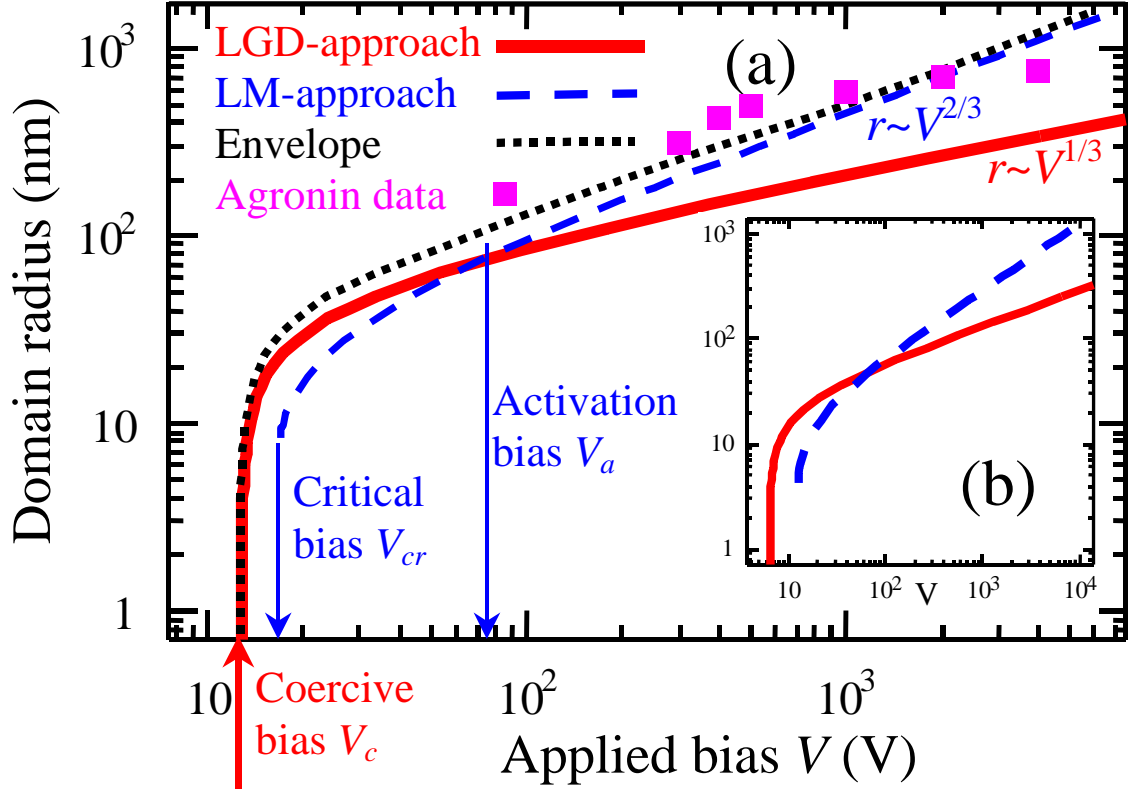

**FIG. 8**. Diagram demonstrating the main features of probe-induced domain formation calculated within LGD-approach (solid curves), LM-approach (dashed curves) and their envelope (dotted curve). LNO material parameters are the same as in Fig.5, $\psi_S = 0.35$ J/m$^2$, probe characteristics $d = 50$ nm (a) and $d = 25$ nm (b). Squares are experimental data reported by Agronin et al [71].

## 6. Summary

    The mechanism of the bias-induced phase transitions and domain formation in the localized electric field of an SPM tip is analyzed using the Landau-Ginzburg-Devonshire approach. This framework allows intrinsic domain wall width and nonlinear correlation effects to be taken into account. The analytical expressions valid for both *first* and *second order* ferroelectrics are derived. The expressions provide insight how the polarization redistribution depends on the gradient energy, nonlinear correlation and depolarization effects, the distribution of the probe's electrostatic potential and the ferroelectric properties of the material.

    The polarization switching is found to proceed in three stages. Below the coercive bias, the polarization distribution inside the subcritical domain nucleus is very smooth, with

polarization maximum directly below the probe. The corresponding electric field distribution is centered in the tip-surface junction area. In contrast, polarization distribution inside the stable domain forming above coercive bias is rectangular-like. Electric field now contains the dipolar component due to the depolarization field induced by the charge domain wall at the tip apex. The corresponding coercive bias for the formation of a stable domain is in reasonable agreement with available experimental results for typical ferroelectric materials. The microscopic origin of the domain elongation in the region where the electric field of the probe is much smaller than the intrinsic coercive field is the positive depolarization field in front of the moving charged domain wall. Domain breakdown through the sample depth occurs for infinitely thin domain walls.

Note that a high PFM response contrast is possible when reversed polarization value near the probe apex is several times higher than the sample spontaneous polarization far from the probe. The condition was obtained with the increase of the applied bias (see Fig.7). This opens a pathway for high-density data storage in ultra-thin layers of ferroelectric materials with high nonlinear field and correlation effects.

**Acknowledgements**

ANM and SVS gratefully acknowledge financial support from National Academy of Science of Ukraine grant N 13-07, joint Russian-Ukrainian grant NASU N 17-Ukr_a (RFBR N 08-02-90434), Ministry of Science and Education of Ukraine grant N GP/F26/042. The research is supported in part (SVK) by the Division of Scientific User Facilities, US DOE, and (PM) Eugene P. Wigner fellowship at ORNL.

**Supplements**

***S.1. Linearized solution in LGD-approach***

Fourier-image on transverse coordinates {$x,y$} of electric field normal component $\tilde{E}_3(\mathbf{k}, z) = -\partial\tilde{\varphi}/\partial z$ calculated from the boundary problem (1) is the sum of external (*e*) and depolarization (*d*) fields [49] is:

$$\tilde{E}_3(\mathbf{k}, z) = \tilde{E}_3^e(\mathbf{k}, z) + \tilde{E}_3^d(\mathbf{k}, z), \tag{S.1a}$$

$$\tilde{E}_3^e(\mathbf{k}, z) = \tilde{V}_e(\mathbf{k}) \frac{\cosh(k(h-z)/\gamma_b)}{\sinh(k h/\gamma_b)} \frac{k}{\gamma_b}, \tag{S.1b}$$

$$\tilde{E}_3^d(\mathbf{k}, z) = \left( \begin{array}{l} \int_0^z dz' \frac{\tilde{P}_3(\mathbf{k}, z')}{\varepsilon_0 \varepsilon_{33}^b} \cosh(k z'/\gamma_b) \frac{\cosh(k(h-z)/\gamma_b)}{\sinh(k h/\gamma_b)} \frac{k}{\gamma_b} + \\ \int_z^h dz' \frac{\tilde{P}_3(\mathbf{k}, z')}{\varepsilon_0 \varepsilon_{33}^b} \cosh(k(h-z')/\gamma_b) \frac{\cosh(k z/\gamma_b)}{\sinh(k h/\gamma_b)} \frac{k}{\gamma_b} - \frac{\tilde{P}_3(\mathbf{k}, z)}{\varepsilon_0 \varepsilon_{33}^b} \end{array} \right). \tag{S.1c}$$

For a semi-infinite sample
$$\tilde{E}_3^d = \left( \begin{array}{l} \int_0^z dz' \frac{\tilde{P}_3(\mathbf{k}, z')}{\varepsilon_0 \varepsilon_{33}^b} \cosh(k z'/\gamma_b) \exp(-k z/\gamma_b) \frac{k}{\gamma_b} + \\ \int_z^\infty dz' \frac{\tilde{P}_3(\mathbf{k}, z')}{\varepsilon_0 \varepsilon_{33}^b} \exp(-k z'/\gamma_b) \cosh(k z/\gamma_b) \frac{k}{\gamma_b} - \frac{\tilde{P}_3(\mathbf{k}, z)}{\varepsilon_0 \varepsilon_{33}^b} \end{array} \right).$$

Here $\gamma_b = \sqrt{\varepsilon_{33}^b / \varepsilon_{11}}$ is the "bare" dielectric anisotropy factor, $\mathbf{k} = \{k_1, k_2\}$ is a spatial wave-vector, its absolute value $k = \sqrt{k_1^2 + k_2^2}$. For a transversally homogeneous media, $\varepsilon_{33}^b = 1$ and static case Eq. (2c) reduces to the expression for depolarization field obtained by Kretschmer and Binder [52].

Perturbation $p$ satisfies linearized equation with appropriate boundary conditions.

$$\left(\alpha + 3\beta P_S^2 + 5\delta P_S^4\right) p - \xi \frac{\partial^2 p}{\partial z^2} - \eta \left(\frac{\partial^2 p}{\partial x^2} + \frac{\partial^2 p}{\partial y^2}\right) = E_3^e + E_3^d[p], \tag{S.2a}$$

$$\frac{\partial p(z=0)}{\partial z} = 0, \qquad \frac{\partial p(z=h)}{\partial z} = 0. \tag{S.2b}$$

Allowing for the radial symmetry of normalized probe potential, $\tilde{w}(k) = d \exp(-kd)/k$, at $h \to \infty$ we obtained the polarization distribution in the form [49]:

$$P_3(\mathbf{r}) = P_S - V \int_0^\infty \frac{k^2}{\gamma_b} dk \cdot J_0\left(k\sqrt{x^2 + y^2}\right) \tilde{w}(k) \frac{(s_2 \exp(-s_1 z) - s_1 \exp(-s_2 z))}{\sqrt{\xi(\eta k^2 + \alpha_S)(s_1^2 - s_2^2)}}. \tag{S.3}$$

Characteristic equation for eigenvalues $s(k)$ is biquadratic, namely $\left(\varepsilon_{33}^b s^2 - \varepsilon_{11} k^2\right)\left(\alpha_S - \left(\xi s^2 - \eta k^2\right)\right) = -\frac{s^2}{\varepsilon_0}$. Renormalized coefficient $\alpha_S = \alpha + 3\beta P_S^2 + 5\delta P_S^4$ is related with spatially averaged generalized susceptibility as $\langle dP_3/dE_e \rangle = 1/\alpha_S$, and thus it should be positive for considered physical situations.

It is seen that for any real values of $k$ values of $s_{1,2}(k)$ are real and the identity is valid:

$$s_1^2 - s_2^2 = \frac{\sqrt{\left(1+\varepsilon_0\left(k^2(\varepsilon_{11}\xi+\eta\varepsilon_{33}^b)+\alpha_s\varepsilon_{33}^b\right)\right)^2 - 4(\varepsilon_0 k)^2 \xi\varepsilon_{11}\varepsilon_{33}^b(\eta k^2+\alpha_s)}}{\varepsilon_0\varepsilon_{33}^b\xi}.$$ For small $k$ values

$$s_1 \approx \sqrt{\frac{1+\alpha_s\varepsilon_0\varepsilon_{33}^b}{\varepsilon_0\varepsilon_{33}^b\xi}} \text{ and } s_2 \approx k\sqrt{\frac{\alpha_s\varepsilon_{11}\varepsilon_0}{1+\alpha_s\varepsilon_0\varepsilon_{33}^b}}.$$

For typical ferroelectric material parameters the inequalities $\varepsilon_0\varepsilon_{33}^b|\alpha_s|\ll 1$, $\sqrt{\varepsilon_0\varepsilon_{33}^b\eta}<1\text{Å}$ and $\sqrt{\varepsilon_0\varepsilon_{33}^b\xi}<1\text{Å}$ are valid, since $\varepsilon_{33}^b\leq 10$. So, parameter $s_1$ value is very high and $\exp(-s_1 z)\ll 1$ once the depth z is more than a lattice constant $a$. Thus, at the sample surface, z=0, and z>>a, Eq. (S.2), can be simplified as:

$$P_3(\mathbf{r}) = -P_S + p(\mathbf{r}),$$

$$p(\rho, z) = p_V \left( \frac{(d+z/\gamma_S)d^2}{\left(L_\perp(d+z/\gamma_S)+(d+z/\gamma_S)^2+\rho^2\right)^{3/2}} + \frac{d^2(d^2+\rho^2)-3d^4}{\gamma_S(d^2+\rho^2)^{5/2}} L_z \exp\left(-\frac{z}{L_z}\right) \right) \approx \frac{p_V(d+z/\gamma_S)d^2}{\left((d+z/\gamma_S)^2+\rho^2\right)^{3/2}}. \quad (S.4)$$

Hereinafter $\rho=\sqrt{x^2+y^2}$ has the meaning of radial coordinate. The length $L_\perp=\sqrt{\eta/\alpha_S}$ originated from the intrinsic width of domain wall, where renormalized coefficient $\alpha_S = \alpha + 3\beta P_S^2 + 5\delta P_S^4$. The correlation length $L_z = \sqrt{\varepsilon_0\varepsilon_{33}^b\xi}$ is extremely small due to the depolarization effects. Effective dielectric anisotropy factor $\gamma_S$, "bare" dielectric anisotropy factor $\gamma_b$ and polarization amplitude $p_V$ are

$$\gamma_S = \sqrt{\frac{\varepsilon_{33}^b}{\varepsilon_{11}} + \frac{1}{\varepsilon_{11}\varepsilon_0(\alpha+3\beta P_S^2+5\delta P_S^4)}} \approx \sqrt{\frac{\varepsilon_{33}}{\varepsilon_{11}}}, \qquad \gamma_b = \sqrt{\frac{\varepsilon_{33}^b}{\varepsilon_{11}}}, \quad (S.5a)$$

$$p_V = \frac{V}{d+L_\perp}\sqrt{\frac{1}{\varepsilon_{11}\varepsilon_0(\alpha+3\beta P_S^2+5\delta P_S^4)}} \approx \varepsilon_{11}\varepsilon_0\frac{V}{d}\sqrt{\gamma_S^2-\gamma_b^2}. \quad (S.5b)$$

When deriving expression (4), we used that the inequalities $2\varepsilon_0\varepsilon_{33}^b|\alpha_S|\ll 1$, $\varepsilon_{33}^b\ll\varepsilon_{33}$, $L_\perp<1$ nm and $L_z<1\text{Å}$ are typically valid for ferroelectric material parameters and

background permittivity $\varepsilon_{33}^b \leq 10$. Hereinafter we use that the inequality $L_z \ll L_\perp \ll d$ is valid. It leads to the approximation $P_3(\rho, z) \approx -P_S + p_V \dfrac{(d + z/\gamma_S)d^2}{\left((d + z/\gamma_S)^2 + \rho^2\right)^{3/2}}$.

Under the reasonable assumption $L_\perp \ll d$, polarization distribution (S.4) produces the following depolarization field:

$$E_3^d(\rho, z) \approx \frac{\gamma_S}{\gamma_S^2 - \gamma_b^2} \frac{d^2 p_V}{\varepsilon_0 \varepsilon_{11}} \left( \frac{(d + z/\gamma_S)}{\gamma_S \left((d + z/\gamma_S)^2 + \rho^2\right)^{3/2}} - \frac{(d + z/\gamma_b)}{\gamma_b \left((d + z/\gamma_b)^2 + \rho^2\right)^{3/2}} \right). \quad (S.6)$$

Depolarization field distribution is shown in Fig.S1.

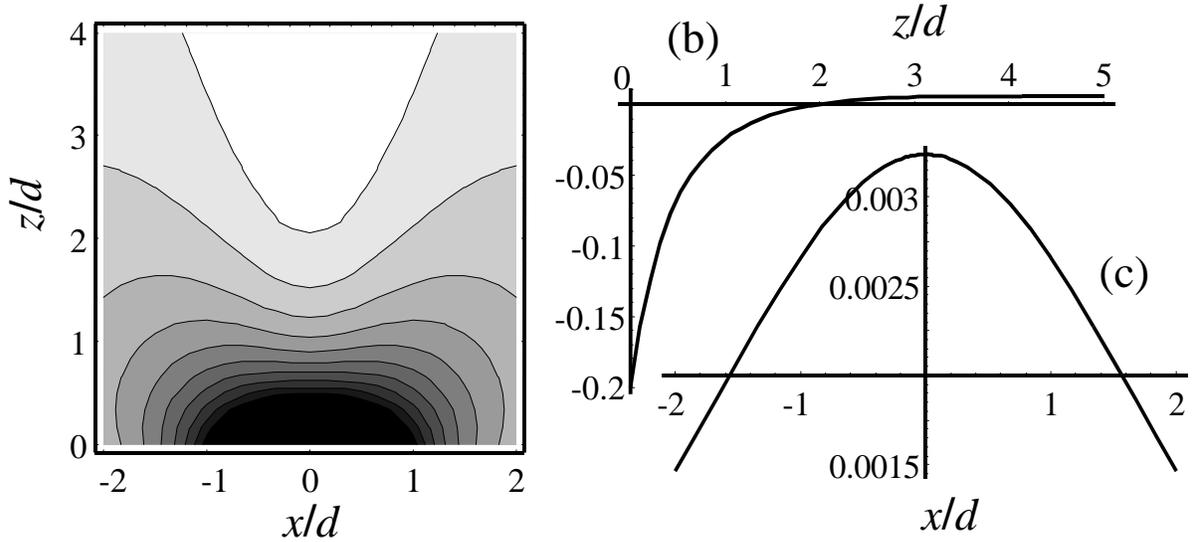

**Fig.S1.** Dimensionless depolarization field distribution $\varepsilon_0 \varepsilon_{33}^b E_3^d / d^2 p_V$ (a), vertical section at $y=x=0$ (b) and cross-version at $y=0$ and $z=10d$ (c). Parameters $\gamma_b=1$ and $\gamma_S=4$.

At $h \to \infty$ the external field is $E_3^e(\rho, z) \approx \dfrac{V(d + z/\gamma_b)d}{\gamma_b \left((d + z/\gamma_b)^2 + \rho^2\right)^{3/2}}$, where $\gamma_b = \sqrt{\varepsilon_{33}^b / \varepsilon_{11}}$.

Allowing for the condition $\alpha_S (\partial P_3 / \partial z) = -\partial^2 \varphi / \partial z^2$, obtained from linearized Eq.(2), and $L_\perp \ll d$, resulting "dressed" electric field, $E_3 = -\partial \varphi / \partial z$, calculated from the problem (1) is

$$E_3(\rho,z) = \frac{V(d+z/\gamma_S)d}{\gamma_S\left((d+z/\gamma_S)^2 + \rho^2\right)^{3/2}}.\tag{S.7}$$

Since the "bare" external field is $E_3^e(\rho,z) = \frac{V(d+z/\gamma_b)d}{\gamma_b\left((d+z/\gamma_b)^2 + \rho^2\right)^{3/2}}$, the identity $E_3(\rho,z) \equiv E_3^d(\rho,z) + E_3^e(\rho,z)$ is valid as expected in linear approximation.

### S.2. Depolarization field in Landauer approximation

In Landauer approximation depolarization field of semi-ellipsoidal domain with sharp domain walls, radius $r$ and length $l$ is:

$$E_W(\zeta(\rho,z),z,r,l) = \frac{(2P_S+\sigma)}{\varepsilon_0\varepsilon_{11}\gamma^2}\begin{cases} -n_D(r,l), & \text{inside the domain} \\ N_D(\zeta(\rho,z),z,r,l), & \text{outside the domain} \end{cases}\tag{S.8}$$

Well-known aspect ratio dependent depolarization factor is introduced as $n_D(r,l) = (r\gamma/l)^2\left((1-(r\gamma/l)^2)^{-3/2}\operatorname{arctanh}\left(\sqrt{1-(r\gamma/l)^2}\right) - (1-(r\gamma/l)^2)^{-1}\right)$, function:

$$N_D(\zeta,r,l) = \frac{(r\gamma/l)^2}{(1-(r\gamma/l)^2)^{3/2}}\left(\begin{array}{l}\sqrt{\frac{l^2-r^2\gamma^2}{l^2+\zeta\gamma^2}} - \operatorname{arctanh}\left(\sqrt{\frac{l^2-r^2\gamma^2}{l^2+\zeta\gamma^2}}\right) + \\ \left(\frac{l^2-r^2\gamma^2}{l^2+\zeta\gamma^2}\right)^{3/2}\frac{z^2(l^2+\zeta\gamma^2)(r^2+\zeta)}{z^2(r^2+\zeta)^2\gamma^2 + \rho^2(l^2+\zeta\gamma^2)^2}\end{array}\right)\tag{S.9}$$

and elliptic coordinate $\zeta = 0.5\left(\sqrt{(r^2-\rho^2+(z^2-l^2)/\gamma^2)^2 + 4\rho^2 z^2/\gamma^2} + \rho^2 - r^2 + (z^2-l^2)/\gamma^2\right)$.

At the domain face $N_D(0,z=0,r,l) = -n_D(r,l)$, while at the domain tip $N_D(0,z=l,r,l) = 1 - n_D(r,l)$.

Electric field at the sample surface should not essentially depend on the counter domain wall width $L_W(l)$, otherwise additional high correlation energy appears. For analytical treatment let us assume that uncompensated bound charge $2P_S$ is continuously distributed between two co-axial semi-ellipsoids with the same aspect ratio: the inner semi-ellipsoid has sizes $\{r,l\}$, the outer one has sizes $\{r+L_W(0), l+L_W(l)\}$. Since their aspect ratio is the same, we obtained that $r/l = (r + L_W(0))/(l + L_W(l))$. Putting $L_W(0) = L_\perp$, we obtained that $L_W(l) \cong L_\perp l/r$.

### S.3. Free energy in LM rigid approach

Within LM thermodynamic approach the nucleus and equilibrium domain sizes are calculated from the free energy excess $G(V,r,l) = G_V(V,r,l) + G_S(r,l) + G_D(r,l)$, where the interaction energy $G_V(r,l) \approx \dfrac{-2\pi d V P_S r^2 l/\gamma}{\left(\sqrt{r^2+d^2}+d\right)\left(\sqrt{r^2+d^2}+d+l/\gamma\right)}$, the domain wall surface energy $G_S(r,l) = \pi \psi_S l r \left( \dfrac{r}{l} + \dfrac{\arcsin\sqrt{1-r^2/l^2}}{\sqrt{1-r^2/l^2}} \right)$ ($\psi_S$ is the surface energy density) and depolarization field energy $G_{DL}(r,l) = \dfrac{4\pi P_S^2 r^2 l}{3\varepsilon_0 \varepsilon_{33}} \dfrac{(r\gamma/l)^2}{1-(r\gamma/l)^2} \left( \dfrac{\text{arcth}\left(\sqrt{1-(r\gamma/l)^2}\right)}{\sqrt{1-(r\gamma/l)^2}} - 1 \right)$ [36]..

### S.4. Lateral growth of cylindrical domain in thin films

Spatial distribution of the electric potential can be represented as

$$\varphi(\rho, z) = Vd \int_0^\infty dk J_0(k\rho) \exp(-kd) \dfrac{\sinh(k(h-z)/\gamma)}{\sinh(kh/\gamma)}. \qquad (S.10)$$

The first term in Eq.(S.10) may be expanded in the image charge series. Under the condition of thick film, $h \gg d$, the series can be cut at the first term, $Vd \Big/ \sqrt{(d+z/\gamma)^2 + \rho^2}$.

Fourier image of electric field $\tilde{E}_3^P(\mathbf{k},z) = \dfrac{Vd}{k} \exp(-kd) \dfrac{\cosh(k(h-z)/\gamma)}{\sinh(kh/\gamma)} \dfrac{k}{\gamma}$ leads to the expression in r-space $E_3(\rho,z) = Vd \int_0^\infty dk J_0(k\rho) \exp(-kd) \dfrac{\cosh(k(h-z)/\gamma)}{\sinh(kh/\gamma)} \dfrac{k}{\gamma}$. Then we obtained

$$E_3(r,0) = Vd \int_0^\infty dk J_0(kr) \exp(-kd) \frac{1+\exp(-2kh/\gamma)}{1-\exp(-2kh/\gamma)} \frac{k}{\gamma} =$$

$$= Vd \int_0^\infty dk J_0(kr) \exp(-kd) \frac{k}{\gamma}\left(1+\exp\left(-\frac{2kh}{\gamma}\right)\right) \sum_{n=0}^\infty \exp\left(-\frac{2knh}{\gamma}\right) = \qquad (S.11)$$

$$= Vd \left( \frac{d}{\gamma(d^2+r^2)^{3/2}} + \sum_{n=1}^\infty \frac{2(d+2hn/\gamma)}{\gamma((d+2hn/\gamma)^2+r^2)^{3/2}} \right)$$